\documentclass[twoside,leqno,twocolumn]{article}
\pdfoutput=1

\usepackage[letterpaper]{geometry}
\usepackage{ltexpprt}

\usepackage{amsmath}
\usepackage{amssymb}
\usepackage{booktabs}
\usepackage{float}
\PassOptionsToPackage{hyphens}{url}\usepackage{hyperref}
\usepackage[square,numbers]{natbib}
\usepackage{todonotes}
\usepackage{listings}
\usepackage{tabularx}
\usepackage{enumitem}
\setlist{nosep}

\begin{document}
\thispagestyle{empty}
\chead{Approved for Public Release; Distribution Unlimited. Public Release Case Number 21-0639}
\lfoot{\copyright 2021 The MITRE Corporation}

\title{\Large Towards Causal Models for Adversary Distractions}
\author{Ron Alford\\
ralford@mitre.org 
\and Andy Applebaum\\
aapplebaum@mitre.org}

\date{}

\maketitle

\begin{abstract} \small\baselineskip=9pt
    Automated adversary emulation is becoming an indispensable tool of network security operators in testing and evaluating their cyber defenses.
    At the same time, it has exposed how quickly adversaries can propagate through the network.
    While research has greatly progressed on quality decoy generation to fool human adversaries, we may need different strategies to slow computer agents.
    In this paper, we show that decoy generation can slow an automated agent's decision process, but that the degree to which it is inhibited is greatly dependent on the types of objects used.
    This points to the need to explicitly evaluate decoy generation and placement strategies against fast moving, automated adversaries.
\end{abstract}

\section{Introduction}

Automated cyber adversaries have the potential to greatly change the speed at which networks are compromised and exploited.
Many of the technologies for slowing and redirecting cyber adversaries revolve around deceiving humans with careful constructed deceits\cite{achleitner2017deceiving,chakraborty2019forge}.
Where human red teams and adversaries are cautious and introspective, current automated adversary techniques are noisy and fast.
In internal testing, MITRE's CALDERA\cite{applebaum2016intelligent} agent can laterally move through a chain of network hosts in under thirty seconds per hop, mostly limited by the windows scheduling technique it exploited.
Given the differences between human-driven and automated attacks, a defender's strategies must account for both.

Numerous works advocate for deceptive emplacements to enhance cybersecurity\cite{stech2016integrating,singer2020shield}.
Almeshekah et al.\cite{almeshekah2014planning} provide a high-level overview of the development and deployment process of deception for cybersecurity, which others have shown increases the frustration and suspicion of human red teams, and presumably, adversaries\cite{ferguson2019tularosa}.
Common network deceptions include honeypot hosts\cite{provos2004virtual}, fake user credentials\cite{herley2008protecting}, and generated content\cite{chakraborty2019forge}, covering both the exploration and exploitation stages of a typical attack. These works, as well as others, make an excellent case for the usage of deception against human operators.

In this work, we focus on how deceptive emplacements on a network can affect an automated adversary's decision making. To prime the discussion, we first provide an overview of the current means of automating decoy placement, from human directed methods to machine learning methods for optimizing interaction with malware.
We then give a brief overview of automated planning, which we use as a proxy for automated cyber agents such as CALDERA.
Using a planning model derived from CALDERA, we show that performance of a state-of-the-art automated planner can be significantly degraded by adding a single new relationship to the environment, but that the effect is highly dependent on the relationship's type.
Finally, we discuss the implications of having different defense strategies against human and automated adversaries, and speculate on ways forward.

\section{Related Work}

Network topology is a common area of contention between deception and counter-deception technologies\cite{al2019autonomous}.
Achleitner et al.\cite{achleitner2017deceiving} give an architecture for creating large numbers of honeypots with believable topological and physical characteristics, using software-defined network (SDN) routers inserted throughout the network. 
Honeypots must be implemented with significant care, as research continues to progress on identifying honeypots, tarpits, and other defensive emplacements\cite{lyon2009nmap,alt2014uncovering}.
While honeypots are expensive to deploy, lighter weight routing policies can obfuscate router layouts, which can prevent adversaries from having the knowledge to launch targeted denial of service attacks\cite{chiang2016acyds,meier2018nethide,trassare2013technique}.

Megatron\cite{ais2020megatron}, an Air Force Research Lab-sponsored project, and the Active Deception Framework\cite{islam2020active} also provide tools for honeypot creation, as well as suite of host and service deception tools.
Neither framework automatically deploy deceptions, and are instead structured around allowing operators to deploy, monitor, and manage deception operations across a network.
Both these systems are refinements over concepts explored in earlier designs for network topology deception\cite{mehresh2012deception,yaman2020autonomous}.

The above systems are potential platforms for intelligent automation of deception emplacement decisions.
Fraunholz and Zimmermann\cite{fraunholz2017towards} consider network deception as a deployment problem, where a set of deception services must be deployed throughout a network in a way that they match their surrounding context (operating systems, services, and IP addresses).
They use machine learning methods to automatically cluster network features, and use these clusters as distributions to generate deception system deployments.
DodgeTron\cite{sajid2020dodgetron} considers the task of malware analysis, and how to create a honey network that maximizes interaction with a malware sample. They use symbolic execution on a malware corpus to catalog system and API calls. The corpus is then clustered by their call traces. New malware is matched against the clusters, and a representative sample from their match is used to select honey things to populate a network. Overall, more than 90 percent of their evaluation samples collected one or more honey things from the network.

Game theory has a wide variety of tools for modeling conflict in uncertain domains. An early application to cybersecurity quantified how increasing deceptions can hit a point of diminishing returns\cite{hespanha2000deception}, after which an attacker switches from informed actions to blind policies (attacking all hosts on a network, for example).
Ferguson-Walter et al.\cite{ferguson2019game} argue that modeling cyber deception as a 2-player sequential hypergame in order allows the defender to capture the uncertainty about the parameters of conflict between adversary and defender\cite{kovach2015hypergame}.
Jajodia et al.\cite{jajodia2017probabilistic} define a probabilistic logic game between attacker and defender around a network scanning use case, where the defender must create fake network scan results to steer an attacker towards hosts which are harder to compromise or less critical to its mission.
To analyze dynamic routing to honeypots, La et al.\cite{la2016deceptive} formulate the problem as a repeated Bayesian game with imperfect information about the type of the attacker, covering the tradeoff between attack detection, attack success, and unnecessary honeypot usage.
Common to these game theoretic approaches is a focus on the role that deception plays on the outcome of the adversary's action.
Our work is complementary, showing that the adversary's reasoning process may be manipulated, slowing it down to provide crucial time for the defenders to respond.

\thispagestyle{empty}
\chead{}
\lfoot{}

\section{Background: Automated Planning}
Automated planning is the study of how to act in an environment to achieve some goal or reward.
Classical planning restricts itself to deterministic, fully-observable, single agent environments with finite, discrete action and state spaces.
Other than full observability, these restrictions suit the execution environment of automated red team agents such as CALDERA\cite{miller2018automated}.

A classical planning problem $\mathcal{P}$ is a tuple $\left(\mathcal{L}, \mathcal{O}, s_0, \mathcal{G}\right)$, defined as follows: 
$\mathcal{L}$ is a finite, first order, function-free logical language which contains \emph{terms} (both variables such as $?x$, $?y$ and constants such as \texttt{computer01}, \texttt{user664}), and \emph{predicate} symbols to denote relationships between terms, e.g., \texttt{(has\_user computer01 user664)}.
We call sentences and predicates from $\mathcal{L}$ \textit{ground} if they contain only constants.
A \textit{state} $s$ in $\mathcal{P}$ is a set of ground predicates from $\mathcal{L}$ which are true in an environment. All ground predicates not in $s$ are assumed to be false (the closed world assumption). The set of all states is $\mathcal{S}$, $s_0\in\mathcal{S}$ is the initial, starting state for the problem.
$\mathcal{G}$ is the goal condition we want to achieve, expressed as a sentence from $\mathcal{L}$.

$\mathcal{O}$ is a set of operators, each defined as a tuple $o=\left(p, e_{add}, e_{del}\right)$, where $p$ is a logical sentence from $\mathcal{L}$ expressing when $o$ can be applied, $e_{add}$ and $e_{del}$ are positive and negative effects of $o$, expressed as sets of predicates which become true or false, respectively. An operator without any free variables in its components is called an \textit{action}. We \textit{ground} an operator $o$ into an action by substituting constants from $\mathcal{L}$ for free variables in $o$.
The set of actions $\mathcal{A}$ is the set of all possible groundings of operators in $\mathcal{O}$. We apply an action whose precondition $p$ is satisfied to a state by deleting its negative effects and adding its positive effects. This forms a state transition function $\gamma: \mathcal{S} \times \mathcal{A} \rightarrow \mathcal{S}$, defined as:
\[
    \gamma\left(s,\left(p, e_{add}, e_{del}\right)\right) =
    \left\{
    \begin{matrix}
        s & \text{ if } s \nvDash p \\
        \left( s  \setminus e_{del} \right) \cup e_{add} & \text{ if } s \vDash p
    \end{matrix}
    \right.
\]
We say a sequence of actions $\left<a_0,a_1,\ldots...,a_n\right>$ is a plan for a problem $\mathcal{P}$ if there are a sequence of states $\left<s_1,\ldots,s_{n+1}\right>$ such that $\gamma(s_i,a_i) = s_{i+1}$ and $s_{n+1} \vDash \mathcal{G}$.
A diverse array of techniques exist for finding plans\cite{ghallab2016automated}.
Solving planning problems can be computationally expensive, EXPSPACE in the worst case for the formalism above\cite{bylander1994computational}.
However, many planning problems are solvable in reasonable time frames\cite{cenamor2019insights}. In particular, cybersecurity planning problems often have monotonic state and action spaces\cite{hoffmann2015simulated}, making them solvable in polynomial time for a fixed domain ($\mathcal{L}$ and $\mathcal{O}$).

\section{Preliminary Results: Predicate Sensitivity}
\label{sec:results}
As preliminary work, we ran a series of experiments looking into the robustness of planners against changes to specific predicates, motivated by the work of \cite{vallati2019robustness}. In this work, the authors seek to understand the robustness of domain-independent planners, running a series of experiments where a hypothetical attacker modifies a planning problem for the purposes of slowing down \emph{any} planner that were to try to solve it, making modifications semi-randomly. Their results are varied, with some planners performing consistently across all of the tests and others showing high variability.


Our work builds on this by moving away from \emph{overall} robustness to instead focus on \emph{targeted} robustness: can we deliberately add knowledge to a planning problem for the \emph{express purpose} of slowing down solvers? This expands the work of \cite{vallati2019robustness} by changing the threat model to an attacker who deliberately tries to slow down the planner, as opposed to one behaving randomly. 

This section outlines the scope and results of some of our initial work, where we look to identify which predicates within a planning problem are most sensitive to modification and cause the most slowdown for a planner. Focusing on cyber, our work considers the case where we are defending against an automated adversary exploiting credential re-use during a post-compromise intrusion. While still preliminary, we believe that these results are important in understanding how deception can be used against future automated adversaries.

\subsection{Cyber as a Planning Problem}


For our tests, we chose to use the ``CALDERA" Planning Domain Definition Language (PDDL) representation \cite{calderaPDDL,miller2018automated}, which was used as part of the 2018 International Planning Competition (IPC) \cite{IPC2018}. This domain features a simple adversary model where we assume the adversary already has a foothold within a Windows enterprise network and is looking to exploit credential re-use to laterally move to each workstation within the network. The domain features 8 actions, each of which corresponds to a typical post-compromise activity such as obtaining domain and host information, dumping credentials from memory, mounting network shares, and copying and executing files remotely for lateral movement. The domain also features several standard object types -- i.e., numbers and strings -- as well as custom types for system components such as users, hosts, domains, etc.

Most important for our experiments are the domain's predicates. In total, the CALDERA domain features 30 predicates -- 3 of which are \emph{status} predicates, which denote the adversary's current status in the environment, and 27 of which are \emph{property} predicates which describe the environment itself. The following are two example property predicates:

{\centering\small\begin{lstlisting}[language=Lisp]
(prop_user ?a - ObservedDomainCredential
           ?b - ObservedDomainUser)
(prop_path ?a - ObservedFile
           ?b - string)
\end{lstlisting}}

The first, \emph{prop\_user}, links a user object to a credential object, while the second links a file to its location.

Of the 27 property predicates, only 14 are controllable: the other 13 predicates used within the CALDERA domain are explicitly controlled by the adversary and thus unavailable as potential decoys. An example here is \emph{prop\_path}, which is set by the adversary when copying a file. Table~\ref{tbl:predicates} lists the 14 controllable predicates and provides a brief explanation of each.

\begin{table*}
    \centering\small
    \begin{tabularx}{\textwidth}{lccX}
    Predicate & Avg. Time (s) & Std. Dev. (s) & Description\\
    \hline
    \emph{mem\_domain\_user\_admins} & 42.3 & 229.2& Denotes a user is an admin on a host. \\
    \emph{prop\_user} & 14.2 & 37.4& Connects a user to a credential. \\
    \emph{prop\_windows\_domain} & 11.1 & 0.9 & Declares the name of a Windows domain.\\
    \emph{mem\_cached\_domain\_creds} & 10.1 & 13.4 & Denotes credentials stored on a given host.\\
    \emph{prop\_sid} & 6.5 & 0.6 & Denotes the security ID of a user.\\
    \emph{prop\_is\_group} & 6.4 & 0.6 & Boolean for if the user is a group.\\
    \emph{prop\_seconds} & 6.4 & 0.6 & Current time's seconds on a given host.\\
    \emph{prop\_dc} & 6.4 & 0.6 & Boolean for if a host is a domain controller.\\
    \emph{prop\_hostname} & 6.4 & 0.6 & String name for a given host.\\
    \emph{prop\_timedelta} & 6.4 & 0.6 & Links a host to a custom timedelta object.\\
    \emph{prop\_fqdn} & 6.3 & 0.6 & String fully qualified domain name for a host.\\
    \emph{prop\_dns\_domain\_name} & 6.3 & 0.5 & DNS domain name for a host.\\
    \emph{prop\_dns\_domain} & 6.3 & 0.5 & Links an object to a Windows domain.\\
    \emph{prop\_username} & 6.2 & 0.6 & String username for a user.\\
    \hline
    \end{tabularx}
    \caption{Predicate sensitivity results}
    \label{tbl:predicates}
    \end{table*}

\subsection{Testing Procedure}
\label{subsec:test}
We use a simple process similar to that of \cite{vallati2019robustness} for evaluating the importance of each predicate. After starting with an initial domain and problem file, we run a baseline planner -- \emph{fast-downward} \cite{helmert2006fast} in default configuration -- to determine initial difficulty. We then iterate as follows:
\begin{enumerate}
\item Select a predicate at random.
\item Select objects at random to match the predicate.
\item Add a new assertion linking the object(s) and predicate.
\item Run the planner on the new problem.
\end{enumerate}
Due to the random nature of this process, we repeat until achieving a set number of trials for each predicate, ultimately recording the average time and standard deviation for each predicate. Note that, by design, we only add \emph{one} observation to the problem file, leaving more complex additions for future work. As an example, we might have two users -- e.g., \emph{john} and \emph{pete} -- who each have their own password; a valid addition would be to link one of the users with the other's password, such as an addition where \emph{john}'s password works for \emph{pete}'s account or vice-versa.




While time consuming, this process offers some advantages as it never creates logically inconsistent problems: by only adding information, there is always guaranteed to be a solution to the problem. However, step 2 can introduce unusual structure (e.g., assigning a user two usernames), but doing so avoids issues where objects are created without enough information (e.g., creating a new user but not assigning that user a username or a password).


\subsection{Results}
We chose to focus on one specific problem within the CALDERA IPC problem set, selecting problem 8 to work with. This problem encodes a network with 6 hosts, 12 users, 3 administrators on each host, and 3 cached credentials on each host. With no modifications, it took fast-downward roughly 6.7 seconds to find a path from the start host to compromise each other host.

Table~\ref{tbl:predicates} shows the results from running our testing procedure in Section~\ref{subsec:test} between 30 and 50 times for each predicate. This table shows that for 10 of the 14 controllable predicates, there was effectively no change observed for a modification -- e.g., linking a second username to a user object with \emph{prop\_username} had no significant impact on planner performance. The other four predicates all showed some slowdown on average, with adding an additional admin to a host having the most slowdown (but highest variability) and adding an additional Windows domain having least variability but not a significant slowdown, with the other two predicates -- associating a credential object with a user and adding additional cached credentials on a host -- falling between.


\subsection{Discussion}
We note three interesting results from our experiments. First, just \emph{adding} information is sufficient for slowing down a planner, even if that information should make it easier for the planner to achieve its objectives (as in the case of additional cached credentials or domain admins).
Additional relationships enable new actions, increasing both the ground problem size and the branching factor of the search space, slowing down the planning process. 

The second interesting feature was on the predicates that were useful for slowing the planner. The 10 predicates that had no impact on the planner were to largely be expected, and three of the significant predicates -- \emph{mem\_domain\_user\_admins}, \emph{prop\_user}, and \emph{mem\_cached\_domain\_creds} -- do not come as a surprise given that modifying each of these fields is a common deceptive maneuver. Less common, however, is the \emph{prop\_windows\_domain} predicate: while it is indeed common to add a new Windows domain, that domain is usually added \emph{substantially} and not merely restricted to just the name of the domain, as in our tests. We believe that this is an important factor specifically for automated adversaries. Most actions require information regarding the Windows domain, and so if there are two domains then the planner needs to investigate actions for each domain when constructing plans. 

The last interesting result from our test is on the variability of modifying the predicates. Adding an extra Windows domain had consistent slowdown with little variability -- examining the domain structure further, we find that the Windows domain \emph{must} be interacted with immediately, and so that action's execution is consistent on all trials where this predicate is modified. The other three significant predicates, however, have much more variability as they can occur in different orders depending on what specifically is being modified -- as an example, adding an additional cached credential on the start host will lead to a very different planning problem than adding an additional cached credential on the final host to be compromised. While this is a fairly intuitive concept -- a decoy's efficacy is correlated to where it lies on the attacker's path -- our results help lend further credence to the notion that understanding the attacker model is critical for composing effective decoys and distractions.

\section{Conclusion \& Future Work}

Where many of the works on defensive cyber deception focus on changing changing adversary actions, our preliminary results point towards an additional defensive goal -- adversary delay.
Our experiments show a significant delay with just a single change to the environment.
By analyzing causal models, network operators may be able to more substantially delay intruders.

With regards to our preliminary results in Section~\ref{sec:results}, we believe there are several ways to extend our work. First, our current approach is time consuming: future work that we plan to conduct will instead look to analytically discover which predicates are most sensitive, as opposed to experientially. Second, our approach only examines one predicate modification at a time. In a real deployment, however, it's likely we can make multiple changes, and that our experiments should therefore consider how sets of predicates interact. Lastly, our preliminary testing methodology only adds predicates linking to existing objects, which at times creates logical inconsistencies; while we should be able to add predicates with existing objects, we also would like to run more tests where we are able to add new objects as well.

\clearpage
\bibliographystyle{siamplain}
\bibliography{didas_ai4cs}

\section*{NOTICE}

Portions of this technical data were produced for the U. S. Government under Contract No. FA8702-19-C-0001 and W56KGU-18-D-0004, and is subject to the Rights in Technical Data-Noncommercial Items Clause DFARS 252.227-7013 (FEB 2014)

\copyright 2021 The MITRE Corporation.
This work is licensed under a \href{https://creativecommons.org/licenses/by-nc-nd/4.0/}{Creative Commons Attribution-NonCommercial-NoDerivatives 4.0 International License}.

\end{document}